\documentclass[10pt,letterpaper,twocolumn]{article}

\usepackage{pr}
\usepackage{hyperref}
\usepackage{amsmath}
\usepackage{subfigure}
\usepackage{epsfig}

\begin{document}

\twocolumn[ 

\title{Negative reflections of electromagnetic waves in chiral media}

\author{Chao Zhang$^{1,2}$ and Tie Jun Cui$^{1*}$}

\address{$^{1}$ Center for Computational Electromagnetics and the
State Key Laboratory of Millimeter Waves, \\ Department of Radio
Engineering, Southeast University, Nanjing 210096, P. R. China.
\\
$^{2} $Dept of Electrical and Computer Engineering, University of
New Mexico, Albuquerque, NM, 87106, USA.}

\begin{abstract}

We investigate the reflection properties of
electromagnetic/optical waves in isotropic chiral media. When the
chiral parameter is strong enough, we show that an unusual
\emph{negative reflection} occurs at the interface of the chiral
medium and a perfectly conducting plane, where the incident wave
and one of reflected eigenwaves lie in the same side of the
boundary normal. Using such a property, we further demonstrate
that such a conducting plane can be used for focusing in the
strong chiral medium. The related equations under paraxial optics
approximation are deduced. In a special case of chiral medium, the
chiral nihility, one of the bi-reflections disappears and only
single reflected eigenwave exists, which goes exactly opposite to
the incident wave. Hence the incident and reflected electric
fields will cancel each other to yield a zero total electric
field. In another word, any electromagnetic waves entering the
chiral nihility with perfectly conducting plane will disappear.

\hskip 1.5mm

\noindent PACS numbers. 78.20.Ci, 41.20.Jb, 42.25.Bs, 42.25.Gy
\end{abstract}

 ] 

The chiral medium was first explored in the beginning of 19th
century for its optical rotation phenomenon. Then it is proved
that right and left-hand circularly polarized waves have different
velocities and hence have different refraction indexes in the
chiral medium. Different polarized rotations correspond to
different modes. Therefore, bi-refraction happens at the boundary
of chiral media even if they are isotropic, due to the coexistence
of two different modes caused by chirality.

In 1968, Veselago introduced the concept of negative refraction
when both permittivity and permeability are simultaneously
negative\cite{Vesalago}. In the currently hot research of
left-handed metamaterials, the chirality may be used to split the
degenerated transverse wave modes. If the chiral parameter is
strong enough or the chirality is combined with electric
plasma\cite{nihility,Pendry04,sailing1,sailing2,
PRE,NJP,PRL,Tretyakov,cheng1}, one eigenwave becomes backward
wave, and a negative refraction is generated naturally in one of
the circularly polarized waves.

The earlier research on chiral media is concentrated in the
negative refraction and the relevant physical properties like the
subwavelength focusing. The task of this paper is to discuss the
extraordinary reflection properties of electromagnetic/optical
waves in isotropic chiral media. We will prove that bi-reflection
exists at the boundary of isotropic media. When the chiral
parameter is strong enough, there will be a \emph{negative
reflection} for one of the reflected eigenwaves. Based on such a
property, we show that a plane mirror instead of a lens can be
used for focusing in the strong chiral medium and get a real image
for paraxial rays. Finally, we discuss the behavior of
electromagnetic waves in a chiral nihility with perfectly
conducting plane. It is proved that only single reflected
eigenwave exists, which goes exactly opposite to the incident
wave. The incident and reflected electric fields will cancel each
other to yield a zero total electric field. Hence we discover an
exotic phenomenon that any electromagnetic waves entering the
chiral nihility with perfectly conducting plane will disappear.

In this paper, we define the right-hand polarized wave is the one
whose electric vector rotates clockwise when looking along the
energy stream. Consider a half-infinite space problem, where the
left region is an isotropic chiral medium and the right region is
a perfectly electric conductor (PEC). An incident right-polarized
wave propagates toward the boundary at an oblique angle
$\theta_{i}$ in the $yoz$ plane, as illustrated in Fig. 1(a).
Here, $k_x=0$ has been assumed under the shown coordinate system.

\begin{figure}[h,t,b]
\centerline{
\includegraphics[width=8cm]{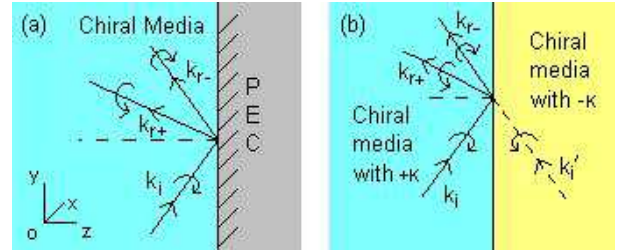}}\caption{\small {Bi-reflection in weak isotropic chiral
medium. (a) Bi-reflection. (b) Image theory.} \label{fig1}}
\end{figure}

Assume that the electric field of incident wave is expressed using
the $e^{-i{\omega}t}$ system as
\begin{equation}\label{eq:eps}
\vec{E}_{i}=\vec{E}_{0} e^{ik_{y}y+ik_{z}z},
\end{equation}
where $\vec{k}=\hat{y}k_{y}+\hat{z}k_{z}$ indicates the
wavevector. According to Maxwell equations and the constitutive
relation for isotropic chiral media\cite{lindell}
\begin{eqnarray}\label{constitu}
\vec{D}={\epsilon}\vec{E}+i{\kappa}\vec{H}, \\
\vec{B}={\mu}\vec{H}-i{\kappa}\vec{E},
\end{eqnarray}
we obtain the following dispersion relation for the wavenumber
$k$:
\begin{equation}
k_{\pm}=\omega(\sqrt{\mu \epsilon}\pm \kappa),
\end{equation}
where ``$+$'' and ``$-$'' represent two different eigenwaves. In
above expressions, $\kappa$ indicates the chirality, which is
assumed to be positive in this paper. Similar dual conclusions can
be easily expanded to the negative chirality.

Correspondingly, the eigenwave vectors are given in terms of the
free variable $E_x$ as
\begin{eqnarray}\label{res}
E_{y} \hskip-2.5mm &=& \hskip-2.5mm {\pm i E_{x} k_{z}}/{k_{\pm}}, \\
E_{z} \hskip-2.5mm &=& \hskip-2.5mm {\mp iE_{x}k_{y}}/{k_{\pm}},\\
H_{x} \hskip-2.5mm &=& \hskip-2.5mm \mp iE_x/\eta, \\
H_{y} \hskip-2.5mm &=& \hskip-2.5mm {E_x k_z}/{k_{\pm}}\eta, \\
H_{z} \hskip-2.5mm &=& \hskip-2.5mm {-E_x k_y}/{k_{\pm}}\eta,
\end{eqnarray}
in which $\eta=\sqrt{{\mu}/{\epsilon}}$ is the wave impedance of
chiral media. It is self-evident that it is left circularly
polarized wave if $k_{y+}^{2}+k_{z+}^{2}=k_{+}^{2}$, and right
circularly polarized wave if $k_{y-}^{2}+k_{z-}^{2}=k_{-}^{2}$.
Hence the incident right polarized wave must match with one of the
eigenwaves and be written as
\begin{equation}
\vec{E}_{i}={E_0}(\hat{x}-\hat{y}{ik_{z-}}/{k_-}+\hat{z}{ik_{y-}}/{k_-})
e^{ik_{y-}y+ik_{z-}z},
\end{equation}
in which $E_0$ is a free variable indicating the amplitude.

First we consider the case of weak chirality, where
$|\kappa|<\sqrt{\mu\epsilon}$. Traditionally, it is regarded as a
natural limit to all chiral media for positive energy requirement.
However, in the recent research on left-handed materials, we know
that the energy calculation in dispersive media is not so simple,
and negative wavenumber do not result in negative energy at
all\cite{Cui}. Hence, it is fairly possible that strong chiral
medium with $\kappa>\sqrt{\mu\epsilon}$ also exists at some
frequency \cite{nihility}, which will be discussed later in this
paper. Under weak chirality, both $k_{+}$ and $k_{-}$ are
positive.

We need point out that reflected waves with different eigenmodes
have different $k$ and $k_z$, but the same $k_y$ due to the phase
matching on the boundary. So we may draw
\begin{equation}
k_{y+}=k_{y-}=k_{y}.
\end{equation}
We assume that the projections of reflected energies and phase
vectors on the $z$ axis are both negative as the common sense.
Though the incident wave is a right-polarized wave, we cannot
ensure whether the reflected wave is right or left polarized.
Hence, we suppose that both exist, and then use the boundary
condition to calculate their coefficients.

It is clear that the projections of right- and left-polarized
reflected wavenumbers on the $z$ axis are $-k_{z-}$ and $-k_{z+}$.
Here, both $k_{z-}$ and $k_{z+}$ are positive for propagating
waves ($k_y<k_{\pm}$). From the boundary condition on the PEC
boundary, we have
\begin{eqnarray}
1+A+B=0, \\
{-ik_{z-}}/{k_-}+A(ik_{z-})/{k_-}-B(ik_{z+})/{k_+}=0,
\end{eqnarray}
where $A$ and $B$ are reflected coefficients of the right- and
left-polarized waves separately. After simple derivation, we
obtain
\begin{eqnarray}\label{AB}
A=({k_{z-}k_{+}
-k_{z+}k_{-}})/({k_{z-}k_+ + k_{z+}k_{-}}), \\
B={-2k_{z-}k_{+}}/({k_{z+}k_- + k_{z-}k_+}).
\end{eqnarray}
Hence, if there is any chirality, we have $k_{z+}\not=k_{z-}$,
leading to $A\not=0$ and $B\not=0$. That is to say, both
circularly polarized reflected waves exist. If there is no
chirality, we have $A=0$. Hence the whole reflected wave is left
polarized for the right-polarized incident wave. And this
degeneration is just the common case of circularly polarized wave
reflection in non-chiral medium.

For the left- and right-polarized reflected waves, their $k$
vectors are: $k_{\pm}=\hat{y}k_y-\hat{z}k_{z\pm}$, and the
corresponding Poynting vectors are written as
\begin{eqnarray}
\Vec{S}_{r+}=\frac{1}{2}\texttt{Re}(\Vec{E}\times
\Vec{H}^{*})=\frac{|E_{x}B|^{2}}{k_{+}}\sqrt{\frac{\epsilon}{\mu}}
(\hat{y}k_y-\hat{z}k_{z+}), \\
\Vec{S}_{r-}=\frac{1}{2}\texttt{Re}(\Vec{E}\times
\Vec{H}^{*})=\frac{|E_{x}A|^{2}}{k_{-}}\sqrt{\frac{\epsilon}{\mu}}
(\hat{y}k_y-\hat{z}k_{z-}).
\end{eqnarray}
We can see that neither $k$ nor $S$ vectors of the two polarized
reflected waves are the same in chiral media.

Hence two different eigenwaves will be generated from the same
incident wave in a boundary shown in Fig. 1(a), resulting in
bi-reflections. It seems not to satisfy geometrical optics
principles at the first glance. However, the chiral medium is a
special material with such unique characters: the refraction
indexes $n$ for a pair of right- and left-polarized waves do not
equal. In other words, the chiral medium is one material to the
right-polarized wave and another to the left one. Considering the
difference between $n_+$ and $n_-$, the direction of each
polarized reflected wave satisfies Fermat principle.

For the right-polarized reflected wave, its reflected angle equals
the incident angle, i.e., $\theta_{r-}=\theta_{i}$, because the
refraction index of reflected wave equals the incident one. For
the left-polarized one, the reflected angle $\theta_{r+}$
satisfies
\begin{equation}\label{snell}
\frac{sin\theta_{i}}{sin\theta_{r+}}=\frac{k_{+}}{k_{-}}=\frac{n_{+}}{n_{-}},
\end{equation}
which is similar to the Snell's law, since the refraction index of
reflected wave is different from the incident one. Here,
$n_{\pm}=\sqrt{\mu\epsilon}\pm\kappa$ represent refraction indexes
of the two eigenwaves in chiral media.

We make another explanation of the bi-reflections in chiral media.
If we consider the PEC boundary as a perfect mirror, then we may
get an image of incident wave as an effective problem shown in
Fig. 1(b), in which the mirrored chiral medium has an opposite
$\kappa$. The chirality is generated by spatial asymmetry, hence
it should be reversed if the material structure is mirrored. In
the mirrored chiral medium, the right-polarized wave corresponds
to $k_+$ and the left-polarized wave to $k_-$. Then we may turn
the reflection problem into a refraction problem. From the
boundary condition, we get the same result as Eqs. (14) and (15),
indicating that bi-refraction happens on the boundary between the
two dual chiral media. Hence we may also put $k_{r\pm}$ as the
transmission of $k'_i$. In other words, bi-reflection shares the
common essence with bi-refraction.

Next, we consider the case of strong chiral media, where more
interesting characters will appear. When
$\kappa>\sqrt{\mu\epsilon}$, we have
$k_-=\sqrt{\mu\epsilon}-\kappa<0$. Hence the right-polarized wave
turns into a backward wave. That is to say, $E, H$ and $k$ form a
left-handed triad and the Poynting vector $S$ is antiparallel to
$k$. However, the left-polarized wave remains right-handed as in
common media. When the incident wave is left circularly polarized,
it is a forward wave, as shown in Fig. 2(a). As the reflection
happens, the left-polarized reflected wave goes normally while the
right-polarized one is a backward wave. Here, we will illustrate
that a \emph{negative reflection} happens for the backward
eigenwave.

\begin{figure}[h,t,b]
\centerline{
\includegraphics[width=7.5cm]{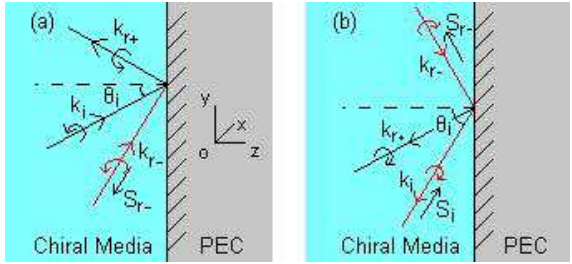}}
\caption{\small {Strong chirality makes negative reflections. (a)
Left-polarized incidence. (b) Right-polarized incidence.}
\label{fig2}}
\end{figure}

Based on the phase matching on the boundary, the $k_y$ components
of both reflected waves should be $+\hat{y}$ directed. Based on
the causality principle, the $S_z$ components of both reflected
waves should be $-\hat{z}$ directed. Hence the left-polarized
reflected wave goes normally as that in the weak-chirality case
with $\theta_{r+}=\theta_{i}$. For the right-polarized reflected
wave, $k_z$ is antiparallel to $S_z$ and hence a \emph{negative
reflection} occurs, where the incident and reflected wavevectors
lie in the same side of the boundary normal, as shown in Fig.
2(a). The reflected angle $\theta_{r-}$ satisfies the Snell-like
law
\begin{equation}\label{snell1}
\frac{sin\theta_{r-}}{sin\theta_{i}}=-\left|\frac{k_{+}}{k_{-}}\right|,
\end{equation}
which yields a negative reflected angle.

Correspondingly, the reflection coefficients are given by
\begin{eqnarray}
A_L=({k_- k_{z+}+k_+ k_{z-}})/({k_- k_{z+}-k_+ k_{z-}}), \\
B_L={-2k_-k_{z+}}/({k_- k_{z+}-k_+ k_{z-}}),
\end{eqnarray}
in which $A_L$ corresponds to the left-polarized wave, and $B_L$
corresponds to the right-polarized wave. For the right-polarized
reflected wave,
\begin{equation}\label{correct}
\vec{E}_{r-}=B_LE_0(\hat{x}-\frac{ik_{z-}}{k_-}\hat{y}+
\frac{ik_y}{k_-}\hat{z})e^{ik_yy+ik_{z-}z},
\end{equation}
\begin{equation}
\vec{S}_{r-}=\frac{|E_xB_L|^2}{k_-}\sqrt{\frac{\epsilon}{\mu}}
(k_y\hat{y}+k_{z-}\hat{z}).
\end{equation}
We remark that $k_{z-}$ has been assigned as positive in this
paper. Considering $k_-<0$, it is clear that $\vec{S}_{r-}$ is
antiparallel to $\vec{k}_{r-}$ for the right-polarized reflected
wave. Negative reflection really happens.

In case of right circularly polarized incident wave, it is a
backward wave, as shown in Fig. 2(b). Now the wave vector for
incident wave is $-k_y\hat{y}-k_{z-}\hat{z}$. Similarly, there are
a normal reflection with $\theta_{r-}=\theta_{i}$, and a negative
reflection for the left-polarized reflected wave with
\begin{equation}\label{snell2}
\frac{sin\theta_{r+}}{sin\theta_{i}}=-\left|\frac{k_{-}}{k_{+}}\right|.
\end{equation}
Correspondingly, the reflection coefficients are given by
\begin{eqnarray}
A_R=({k_+k_{z-}+k_-k_{z+}})/({k_+k_{z-}-k_-k_{z+}}), \\
B_R={-2k_+k_{z-}}/({k_+k_{z-}-k_-k_{z+}}),
\end{eqnarray}
in which $A_R$ corresponds to the right-polarized wave, and $B_R$
corresponds to the left-polarized wave. For the left-polarized
reflected wave,
\begin{equation}
\vec{E}_{r+}=B_RE_0(\hat{x}-
\frac{ik_{z+}}{k_+}\hat{y}+\frac{ik_y}{k_+}\hat{z})e^{-ik_yy-ik_{z+}z},
\end{equation}
\begin{equation}
\vec{S}_{r+}=\frac{|E_xB_R|^2}{k_+}\sqrt{\frac{\epsilon}{\mu}}
(-k_y\hat{y}-k_{z+}\hat{z}).
\end{equation}
We see that the left-polarized reflected wave and the incident
wave lie in the same side of normal. Negative reflection happens
again.

Using such unusual reflection properties, we may realize partial
focusing of a source using a simple PEC mirror. Actually the field
generated by a source can be decomposed as left- and
right-polarized waves. For both polarized-wave incidences, the
reflected waves will be partially focused, as shown in Figs. 3(a)
and 3(b), respectively.

\begin{figure}[h,t,b]
\centerline{
\includegraphics[width=8cm]{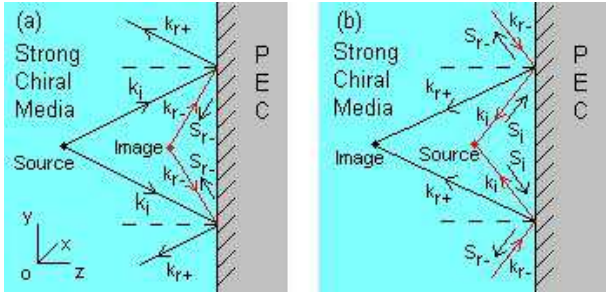}}
\caption{\small {Partial focusing of a source due to the negative
reflections. (a) Left-polarized incidence. (b) Right-polarized
incidence.} \label{fig3}}
\end{figure}

Taking paraxial approximation in Gaussian optics, we may get a
reasonably good image for this partial focusing. Assume that the
distance from the source to PEC mirror is $s$. Then the position
of image point is $s'_L=-sk_-/k_+$ for the left-polarized incident
wave, and $s'_R=-sk_+/k_-$ for the right-polarized incident wave.

It is true that the positively reflected wave will diverge, not
participating in the partial focusing but forming an imaginary
image. However, in the paraxial case, we have $k_+\cong{k_{z+}}$
and $k_-\cong{-k_{z-}}$ for the strong chiral medium. Hence the
amplitudes of positively-reflected waves are close to zero, which
may be neglected. That is to say, most paraxial rays reflected
negatively for partial focusing.

Considering the negative reflections in the strong chiral medium,
some conclusions in the conventional Gaussian optics need to be
improved. For one thing, the real images we get are not upsidedown
as real images always do. On the other hand, we may generalize our
analysis into spherical reflection surface. In strong chiral
medium, the reflection relationship between object and image
distances can be written as:
\begin{eqnarray}\label{gauss}
\frac{k_+}{s}+\frac{k_-}{s'_L}=-\frac{k_++k_-}{R},
\\\label{gauss2}
\frac{k_-}{s}+\frac{k_+}{s'_R}=-\frac{k_++k_-}{R},
\end{eqnarray}
where $R$ is the radius of the spherical surface, which is
positive if convex and negative if concave, and $s'<0$ for the
imaginary image. In the weak chirality case, we may draw the same
results as those in Eqs. (\ref{gauss}) and (\ref{gauss2}) though
there is no negative reflection. These are the general reflection
relationships of all chiral media in paraxial Gaussian optics.

It will be more interesting to discuss a special case of chiral
medium: the chiral nihility with $\mu\epsilon=0$\cite{nihility}.
In chiral nihility, we easily have $k_\pm=\pm\omega\kappa$. Hence
the corresponding physical features are quite similar to those in
the strong-chirality medium, and the formulations (19)-(28) can be
directly used. For propagating waves ($|k_y|<\omega\kappa$), we
obtain $k_{z-}=k_{z+}$ under our definition in this paper.

If the incident wave is left polarized, one easily obtains $A_L=0$
and $B_L=-1$. That is to say, the left-polarized reflected wave
disappears, and a total reflection occurs to the right-polarized
reflected wave, as shown in Fig. 4(a). Here, the wavevectors of
incident and reflected waves are the same (directing to the
up-right direction), while the Poynting vectors are opposite.

\begin{figure}[h,t,b]
\centerline{
\includegraphics[width=8cm]{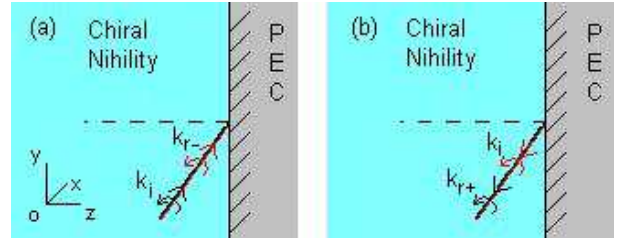}}
\caption{\small {The reflections in chiral nihility. (a)
Left-polarized incidence. (b) Right-polarized incidence.}
\label{fig4}}
\end{figure}

When the incident wave is right polarized, we then have $A_R=0$
and $B_R=-1$. That is to say, the right-polarized reflected wave
disappears, and a total reflection occurs to the left-polarized
reflected wave, as illustrated in Fig. 4(b). Again, the
wavevectors of incident and reflected waves are the same
(directing to the down-left direction), while the Poynting vectors
are opposite.

Based on the above discussions, we can easily show that the
totally reflected electric fields counteract the incident electric
fields exactly in both polarized incidences in the chiral
nihility, which results in zero total electric fields. If
$\epsilon=0$ and $\mu\neq 0$, we can show that all magnetic fields
must be zero from the dispersion equation. In such a case, all
total electric and magnetic fields disappear in the chiral
nihility. If $\epsilon=0$ and $\mu=0$, the magnetic fields may
exist because the electric and magnetic fields are decoupled
completely.

For evanescent waves ($|k_y|>\omega\kappa$), however, we have to
set $k_{z-}=-k_{z+}$ to satisfy the causality under our definition
in this paper. If the incident wave is left polarized, then
$A_L\to\infty$ and $B_L\to\infty$, which is similar to the case of
Pendry's perfect lens\cite{perfectlens}.

It is more interesting to consider a chiral nihility bounded by
two PEC mirrors. When a wave is excited in the chiral nihility,
the wave will be totally reflected forwardly and backwardly again
and again between two mirrors based on the earlier discussions, as
shown in Fig. 5. All waves from the source will focus at the
source point. Using the boundary conditions, we have shown exactly
that the total reflected electric fields including left- and
right-polarized components at any points satisfy
\begin{equation}
\vec{E}_{r}=\vec{E}_{r+}+\vec{E}_{r-}=-\vec{E}_{i},
\end{equation}
which is valid to both \emph{propagating-wave and evanescent-wave}
incidences. Hence, the total electric fields at any points inside
the chiral nihility are zero. If $\epsilon=0$ and $\mu\neq 0$, the
total magnetic fields are also zero. Hence, a source could not
radiate effectively inside the chiral nihility bounded by two PEC
mirrors.

\begin{figure}[h,t,b]
\centerline{
\includegraphics[width=6cm]{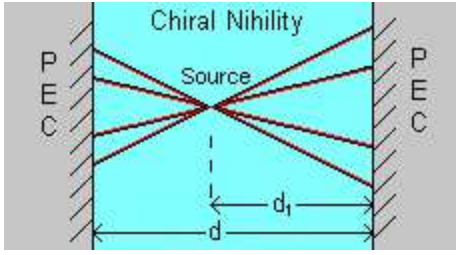}}
\caption{\small {Chiral nihility bounded by two PEC mirrors.}
\label{fig5}}
\end{figure}

In conclusions, negative reflections occur at the boundary of
strong-chiral medium and PEC mirror, which directly result in
partial focusing using a simple plane mirror. Any propagating
waves entering the chiral nihility ($\epsilon=0$ and $\mu\neq 0$)
with a PEC plane will disappear. Any sources could not radiate
inside the chiral nihility bounded by two PEC mirrors.

This work was supported in part by the National Basic Research
Program (973) of China under Grant No. 2004CB719802, in part by
the National Science Foundation of China for Distinguished Young
Scholars under Grant No. 60225001, and in part by the National
Doctoral Foundation of China under Grant No. 20040286010. Email:
tjcui@seu.edu.cn.

\end{document}